\documentclass{isprs} 
\usepackage{graphicx}
\usepackage{url}
\usepackage{setspace}
\usepackage{geometry} 
\usepackage{epstopdf}
\usepackage[labelsep=period]{caption}  
\usepackage[british]{babel} 
\usepackage[hang]{footmisc}
\usepackage{float}
\usepackage{subfigure}
\usepackage{subcaption}
\usepackage{algorithm} 
\usepackage{algpseudocode}
\usepackage{graphicx}
\usepackage{changepage}
\usepackage{booktabs}
\usepackage{wrapfig}
\usepackage{url}
\usepackage{lineno}
\usepackage{xcolor}

\emergencystretch=\maxdimen
\begin{document}

\title{Filling holes in LoD2 building models}
\date{}


\author{
Weixiao Gao\textsuperscript{1}\thanks{Corresponding author}, Ravi Peters\textsuperscript{2}, Hugo Ledoux\textsuperscript{1}, Jantien Stoter\textsuperscript{1} }



\address{
	\textsuperscript{1 }Delft University of Technology, The Netherlands - (w.gao-1,h.ledoux,j.e.stoter)@tudelft.nl\\
	\textsuperscript{2 }3DGI, Zoetermeer, The Netherlands - ravi.peters@3dgi.nl\\
}




\abstract{

This paper presents a new algorithm for filling holes in Level of Detail 2 (LoD2) building mesh models, addressing the challenges posed by geometric inaccuracies and topological errors. Unlike traditional methods that often alter the original geometric structure or impose stringent input requirements, our approach preserves the integrity of the original model while effectively managing a range of topological errors. The algorithm operates in three distinct phases: (1) pre-processing, which addresses topological errors and identifies pseudo-holes; (2) detecting and extracting complete border rings of holes; and (3) remeshing, aimed at reconstructing the complete geometric surface. Our method demonstrates superior performance compared to related work in filling holes in building mesh models, achieving both uniform local geometry around the holes and structural completeness. Comparative experiments with established methods demonstrate our algorithm's effectiveness in delivering more complete and geometrically consistent hole-filling results, albeit with a slight trade-off in efficiency. The paper also identifies challenges in handling certain complex scenarios and outlines future directions for research, including the pursuit of a comprehensive repair goal for LoD2 models to achieve watertight 2-manifold models with correctly oriented normals. Our source code is available at \url{https://github.com/tudelft3d/Automatic-Repair-of-LoD2-Building-Models.git}. 
}


\maketitle


\section{Introduction}\label{sec:Introduction}
As geographic information system and computer graphics technologies continue to advance, an increasing number of sophisticated 3D city models are being generated~\cite{dollner2006virtual,Helsinki3d,peters2022automated}. These models play a crucial role in enhancing urban visualization~\cite{mao2011online,zhang2014web}, aiding urban planning~\cite{murata20043d,Chen2011}, and providing a robust platform for analyzing essential urban data~\cite{geiger20133d}. The building models in these 3D city models are particularly significant, as they capture the fundamental aspects of a city's planning, construction, and development~\cite{biljecki2015applications}.

In the realm of 3D building models, Building Information Modeling (BIM) is often employed for detailed building management~\cite{kolbe2021semantic}. BIM models are characterized by their intricate semantic and geometric details, offering a high degree of precision and information. Concurrently, the construction of building models in large scale 3D urban environments typically adheres to the standards set forth by CityGML models~\cite{kolbe2005citygml}. The Level of Detail 2 (LoD2) building model is notably prevalent in this area. LoD2 models provide a more comprehensive geometric representation of buildings, particularly in terms of roof structures, when compared to their LoD1 counterparts which have only flat horizontal roofs. Furthermore, LoD2 models strike a balance between detail and ease of construction, making them a more practical choice for large-scale urban building modeling than the more complex LoD3 models.

\begin{figure}[ht!]
\centering
\subfigure[]{\label{fig:fig1a}\includegraphics[width=0.48\columnwidth]{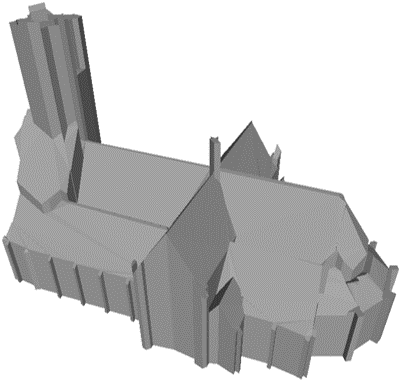}}
\subfigure[]{\label{fig:fig1b}\includegraphics[width=0.48\columnwidth]{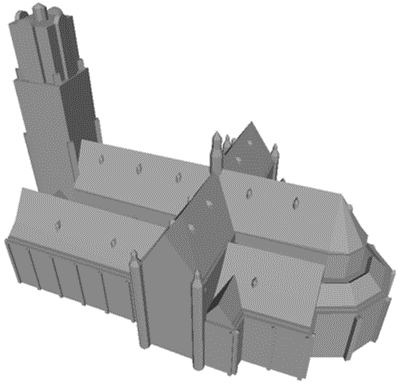}}
\caption{(a) displays a Lod2 building model reconstructed automatically, while (b) presents a manually reconstructed LoD2 building model.}
\label{fig:fig1}
\end{figure}

LoD2 building models are depicted as polygon soups or triangular meshes, created via automatic or (semi-)manual methods. Automatic modeling relies on input data quality for accuracy~\cite{peters2022automated} and aims for watertight, manifold outcomes with geometric constraints~\cite{huang2022city3d}. However, inaccuracies may arise from incomplete data, as shown in Figure~\ref{fig:fig1a}. In contrast, (semi-)manual modeling, integrating GIS, imagery, and elevation, offers higher geometric accuracy (see Figure~\ref{fig:fig1b}). Yet, software limitations, modeler expertise, or data conversion errors can introduce geometric and topological errors, including duplicates, non-manifolds, and holes~\cite{wagner2012geometric}. For applications like fluid simulations, energy estimation, and geometric analyses, error-free, watertight 3D models are crucial~\cite{ledoux2013validation,biljecki2016most}. This research focuses on hole completion in LoD2 building models.

A significant number of LoD2 city models are currently plagued with varying degrees of geometric and topological errors~\cite{wagner2012geometric,biljecki2016most}. Addressing these issues manually would be an arduous and resource-intensive task. Consequently, there is a pressing demand for automated methods capable of both detecting and precisely rectifying these errors. Prior research has yielded encouraging outcomes in identifying anomalies in 3D urban models~\cite{ledoux2013validation,ledoux2018val3dity}. In terms of repairing building models, existing research predominantly falls into two categories: reconstruction-based repair and topological-based local repair. 

\begin{figure}[ht!]
\begin{center}
        \includegraphics[width=1.0\columnwidth]{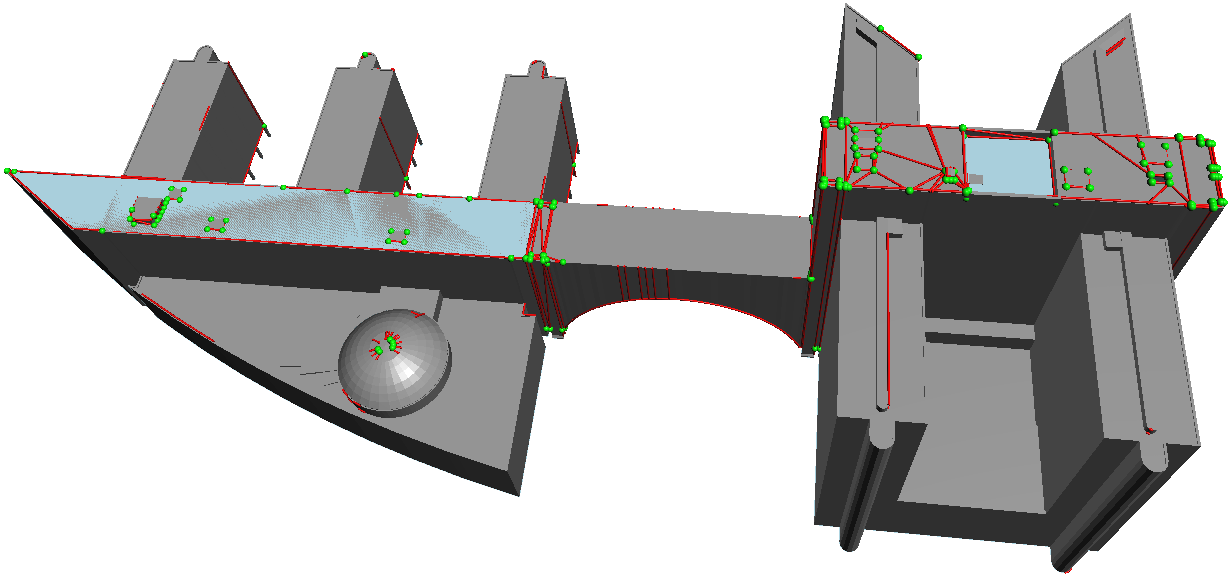}
	\caption{The LoD2 building mesh exhibits several errors, including non-manifold vertices highlighted in green, faces with incorrect normals indicated by the back faces of the mesh depicted in light blue, and boundaries of holes marked in red.}
\label{fig:fig2}
\end{center}
\end{figure}

Reconstruction-based methods can ensure watertight models to a degree, but often alter the original geometry and rely on preconditions like correct normals or semantic labels, posing challenges for inherently deficient models~\cite{sindram2016voluminator,zhao2018hsw,yu2022repairing}. Our approach avoids such strict input requirements and instead focuses on hole detection and repair within the model, preserving geometric integrity. Conversely, traditional topological-based local repairs, arising from computer graphics research, assume 2-manifold inputs with correct normals, utilizing half-edge data structures for hole boundary detection and local geometry for triangulation~\cite{liepa2003filling,zhao2007robust}. However, urban building models, which may have non-manifolds, duplicate vertices, or misoriented normals~\cite{wagner2012geometric} as shown in Figure~\ref{fig:fig2}, present difficulties in accurately identifying holes topologically, leading to potential misrepairs. These conventional methods are thus not universally applicable to all LoD2 building models.

To address this gap, this paper introduces an innovative approach leveraging the continuity of mesh edges and the formation of virtual faces. This method effectively detects and fills holes in LoD2 building models without imposing stringent requirements on the input data, offering a versatile and adaptable solution for urban model repair.

\section{Related work}\label{sec:related_work}
In reviewing existing literature, we have identified two primary categories of urban model repair techniques: reconstruction-based repair and topological-based local repair. Our research is particularly focused on strategies that effectively address the completion of holes in mesh models.

\subsection{Reconstruction-based repair}\label{sec:recons_repair}

Reconstruction-based repair of mesh models typically involves using flawed models as input. And it is aiming to rebuild a complete mesh model by creating voxels or polyhedron to extract surface details and topological structures. These methods have their merits in restoring some missing or incorrect geometric structures, but they are not without limitations. The voxel-based approach~\cite{mulder2015automatic,sindram2016voluminator}, for instance, is limited by factors such as resolution, accuracy, computational demands, and a tendency to either introduce extraneous geometric structures or omit fine details~\cite{mulder2015automatic}. Polyhedron-based methods often stipulate specific conditions for the input model, such as continuous normal vectors, correct mesh topology~\cite{zhao2018hsw}, while allowing for the absence of certain geometric structures (like parts of a wall), and the inclusion of semantic labels~\cite{yu2022repairing}. Moreover, the reconstruction process may inadvertently add unnecessary geometric elements. Although capable of filling in missing information, these methods may alter the model's original geometric structure, either by introducing redundant elements that were not present originally or by omitting crucial components. Our proposed method, however, directly targets the areas with holes in building models without modifying the original structure and imposes no stringent requirements on the input model.

\subsection{Topological-based local repair}\label{sec:topo_repair}

Local repair methods that leverage mesh topology characteristics typically employ the half-edge data structure for detecting edges of holes~\cite{liepa2003filling,zhao2007robust}. They then reconstruct the triangular mesh surface of the holes based on the provided geometric features of these edges. Subsequent post-processing, which includes smoothing and vertex position optimization, is applied to achieve uniformity in the local geometry. Nevertheless, these techniques necessitate that the input mesh is 2-manifold and struggle with larger holes. Moreover, the reconstruction of sharp geometric features is limited only to the local regions at the edges of the holes. In contrast, the method we introduce in this paper is adept at managing non-manifold mesh models and larger hole areas, enabling the restoration of sharp features throughout the entirety of the holes.

\section{Methodology}\label{sec:method}
\begin{figure}[ht!]
\begin{center}
        \includegraphics[width=0.9\columnwidth]{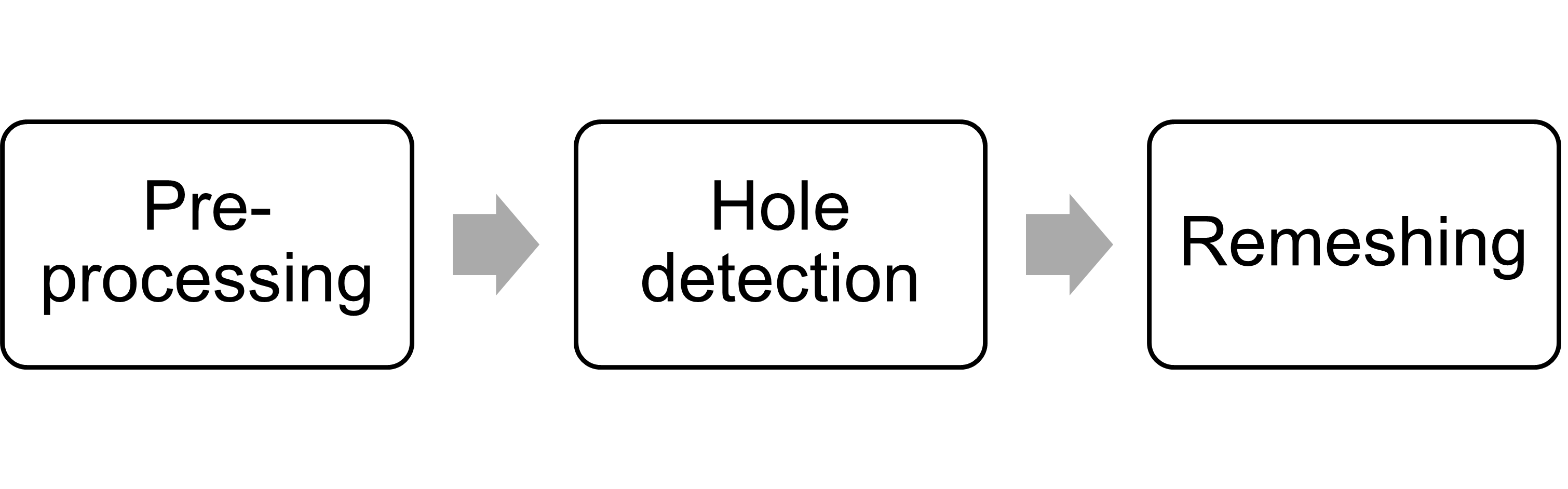}
	\caption{Pipeline for hole filling in building mesh models.}
    \label{fig:fig3}
\end{center}
\end{figure}

Our aim is to fill holes in LoD2 building mesh models. It should be emphasized that the holes being repaired here specifically refer to geometric holes in the model, rather than topological holes or discontinuous gaps. The input for our process comprises piecewise planar LoD2 building mesh models. We approach the input data without pre-established conditions or limitations regarding its geometric attributes, topological characteristics, or semantic information. Upon identifying data with holes, we initiate a comprehensive hole-filling procedure. As shown in Figure~\ref{fig:fig3}, our methodology unfolds in three phases: pre-processing, hole detection, and remeshing.

\subsection{Pre-processing}\label{sec:pre_process}
The objective of pre-processing is to conduct a targeted repair on triangle meshes exhibiting topological errors, and to systematically identify and mark vertices and edges with topological inaccuracies. This preparation is crucial for the efficient detection and handling of holes in subsequent stages.

The process initiates with the identification and resolution of self-intersecting faces within the triangle mesh. We employ remeshing in these areas to rectify the half-edge data structure. Importantly, as the newly generated vertices during remeshing are confined within the planes of the triangles, the local geometric structure of the mesh remains unaltered.

\begin{figure}[ht!]
\begin{center}
        \includegraphics[page=1,width=0.32\columnwidth]{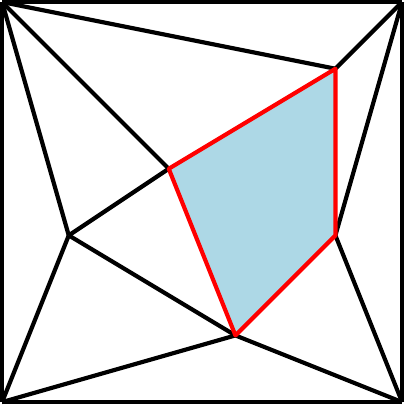}
        \hfill
        \includegraphics[page=2,width=0.32\columnwidth]{figures/fig4.pdf}
        \hfill
        \includegraphics[page=3,width=0.32\columnwidth]{figures/fig4.pdf}
        \caption{A 2D illustration depicts true and pseudo-holes in a 3D mesh. The left figure shows a true hole, while the middle and right display pseudo-holes. Red edges outline the hole's boundaries, and the light blue area indicates the real gap. Vertices in green highlight duplicate presence.}
\label{fig:fig4}
\end{center}
\end{figure}

Subsequently, we concentrate on addressing invalid pseudo-holes with tightly attached boundaries through a detailed stitching operation. Pseudo-holes, which are openings that resemble gaps without signifying real separations in the mesh, typically emerge from issues such as duplicated vertices, overlapping edges, and non-manifold edges. 
In contrast, true holes represent valid gaps with actual separations. Both true and pseudo-holes are geometric holes. Our current focus is specifically on pseudo-holes resulting from duplicated vertices (see Figure~\ref{fig:fig4}). The stitching process, leveraging the half-edge data structure, involves aligning and merging the edges where the source and target vertices of one half-edge (h1) correspond with the target and source vertices of another (h2).

The final step involves the detection and categorization of duplicate vertices and overlapping edges. This identification is critical for the effective management of these anomalies during hole detection. To address duplicate vertices, we construct a kd-tree encompassing all mesh vertices, which allows us to efficiently locate and mark close vertices within a specified threshold range. In the case of overlapping edges, our aim is to pinpoint and mark three specific edges: degenerate edges, edges that share the same endpoints, and edges that, while being collinear, have distinct endpoints.

\begin{figure}[ht!]
\centering
\subfigure[Input]{\label{fig:fig5a}\includegraphics[width=0.48\columnwidth]{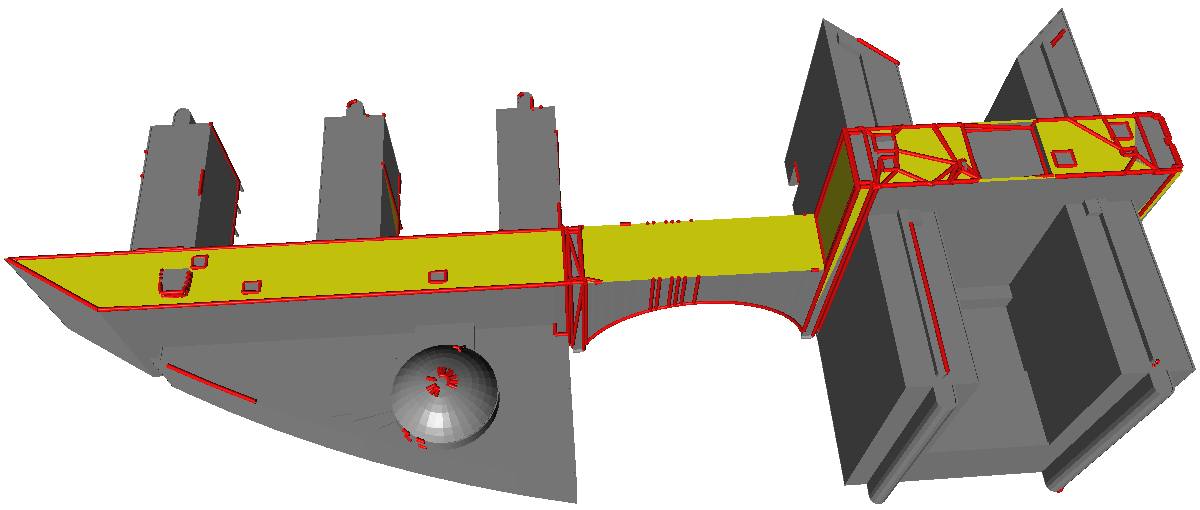}}
\subfigure[Output]{\label{fig:fig5b}\includegraphics[width=0.48\columnwidth]{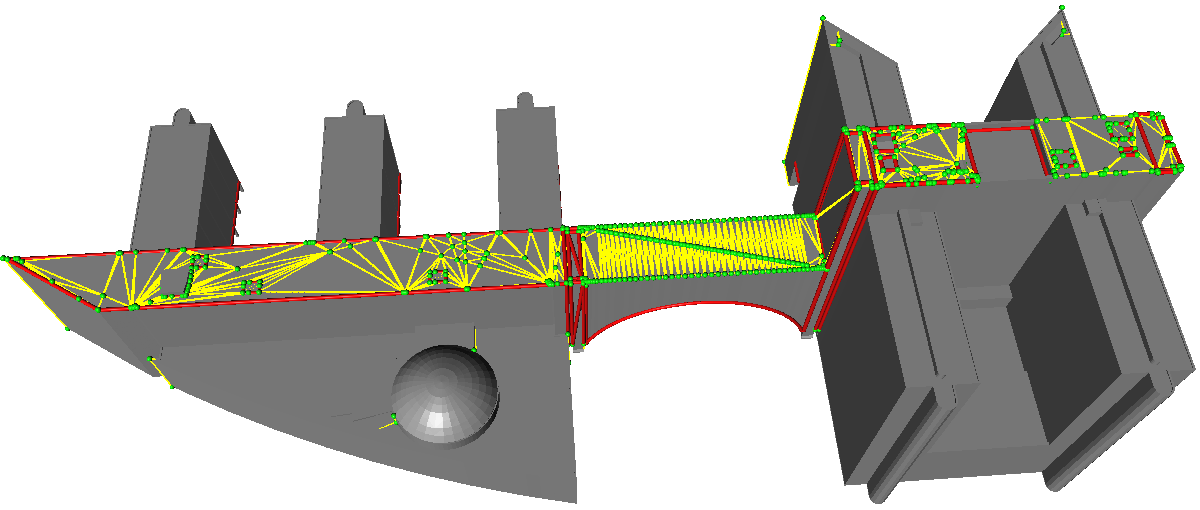}}
\caption{The left figure illustrates the input mesh with self-intersecting faces in yellow and the boundaries of true and pseudo-holes in red. The right figure displays the output mesh, highlighting duplicated vertices in green, duplicated edges in yellow, and maintaining the remaining hole boundaries in red.}
\label{fig:fig5}
\end{figure}

Through this pre-processing phase, we effectively eliminate errors arising from self-intersecting faces and closely attached borders of pseudo-holes (see Figure~\ref{fig:fig5a}). Additionally, we mark vertices and edges that exhibit duplication within the output mesh data (see Figure~\ref{fig:fig5b}). This preparatory work lays the groundwork by providing essential topological insights, setting the stage for the forthcoming hole detection process.

\begin{figure*}[ht!]
\centering
\subfigure[Input]{\label{fig:fig6a}\includegraphics[width=0.24\textwidth]{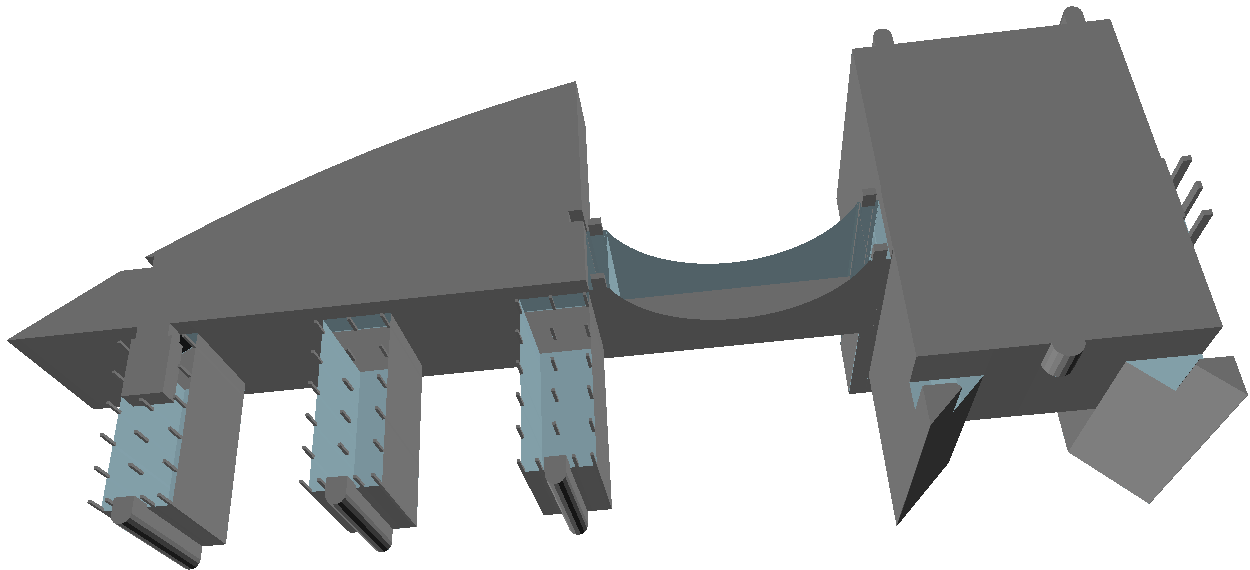}}
\subfigure[All holes]{\label{fig:fig6b}\includegraphics[width=0.24\textwidth]{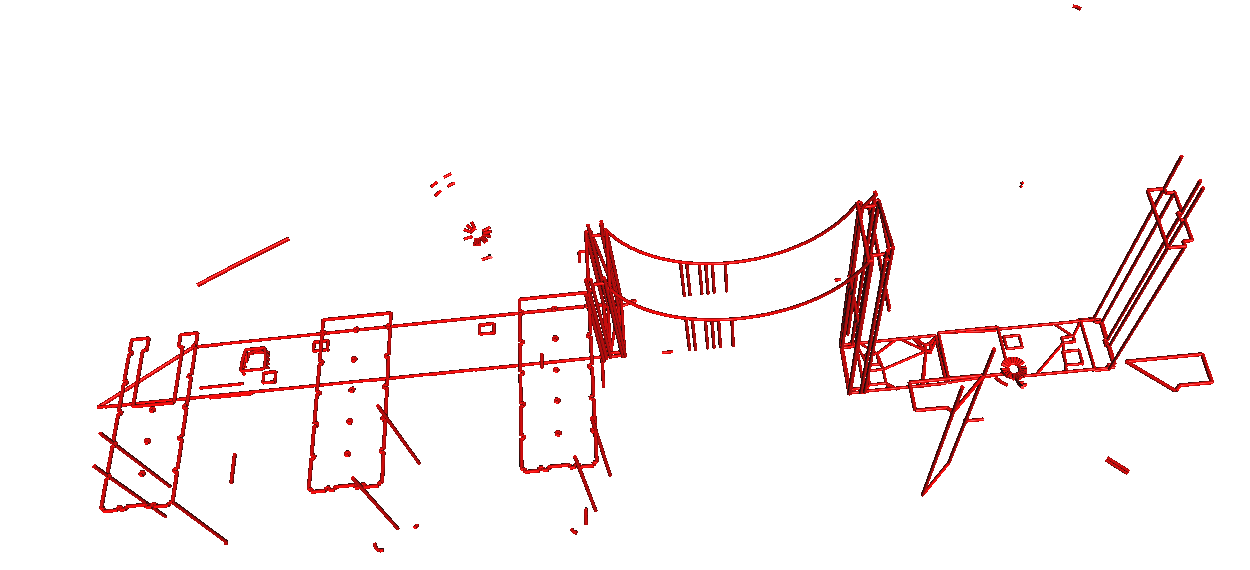}}
\subfigure[True holes]{\label{fig:fig6c}\includegraphics[width=0.24\textwidth]{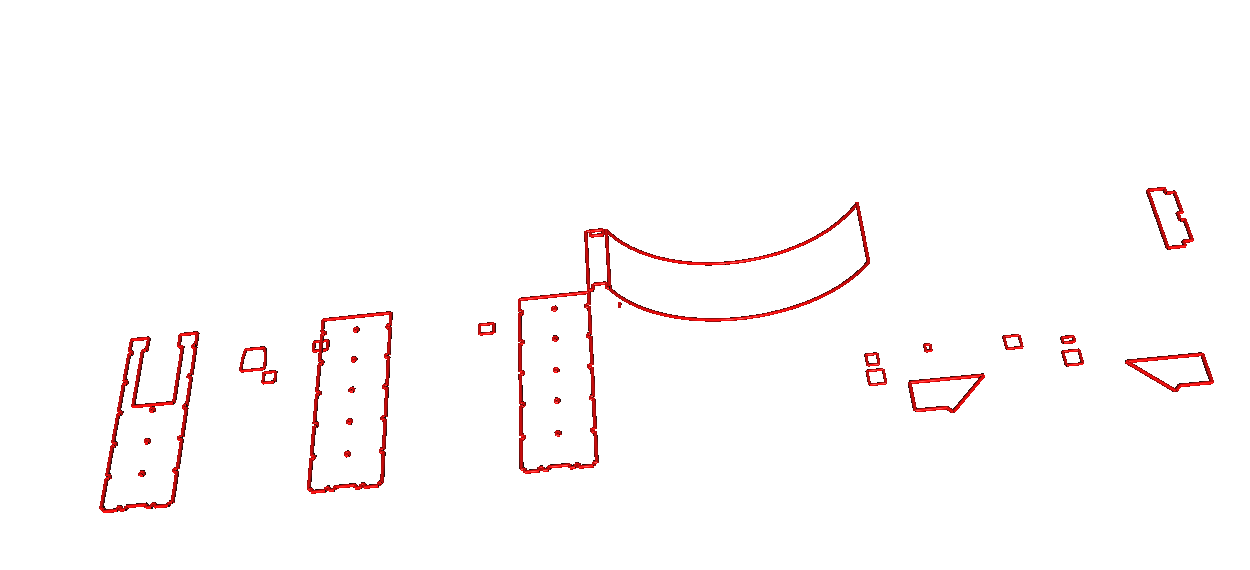}}
\subfigure[Remeshed holes]{\label{fig:fig6d}\includegraphics[width=0.24\textwidth]{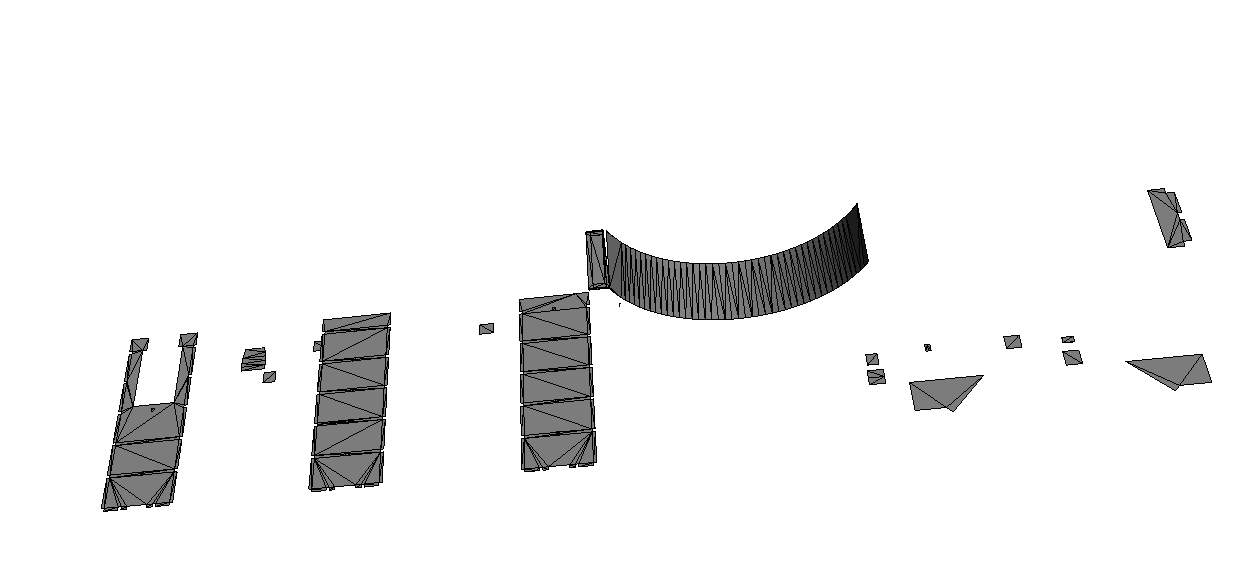}}
\caption{Visualization of intermediate stages in the proposed hole filling process.}
\label{fig:fig6}
\end{figure*}

\subsection{Hole detection}\label{sec:hoel_detect}

The objective of this step is to accurately identify all true holes, distinguishing them from pseudo-holes arising due to overlapping or non-manifold edges (see Figure~\ref{fig:fig6a} to ~\ref{fig:fig6c}). To accomplish this, we assemble triangles using adjacent half-edges and conduct intersection tests with every incident face of their edge vertices. This process is essential for choosing adjacent half-edges that pass intersection tests as candidates for true holes. Subsequently, we complement these selected candidates and extract their complete, closed border rings.

\begin{algorithm}
\caption{Hole Detection Algorithm}
\label{alg:alg1}
\begin{algorithmic}[1]
\For{each halfedge $h$ in mesh}
    \If{$h$ is not visited and is a valid border halfedge}
        \State Mark $h$ as visited and use $h$ as seed halfedge $h_s$
        \Repeat
            \If{preceding halfedge $h_p$ exists and valid}
                \State Collect all incident facets from $h$ and $h_p$ vertices
                \State Create a triangle $\alpha$ from $h$ and $h_p$
                \For{each face $\beta$ in all incident facets}
                    \State Compute maximum distance $D$ from triangle $\alpha$ to $\beta$ (Eq: \ref{equ:1})
                    \State Compute area ratio $A$ from area $A_{\beta}$ of $\beta$ and intersected area $A_w$ between projected $\alpha$ and $\beta$ (Eq: \ref{equ:2})
                    \If{$D < \epsilon_d$ and $\frac{A_w}{A_{\beta}} > \epsilon_t$}
                        \State $\alpha$ intersects with $\beta$
                        \State \textbf{break}
                    \Else
                        \State $\alpha$ does not intersect with $\beta$
                    \EndIf
                \EndFor
                \If{$\alpha$ does not intersect with any $\beta$}
                    \State Collect $h_p$ and $h$ to current border ring
                \EndIf
            \EndIf
            \State Move $h$ to next halfedge
        \Until{$h$ is not equal to seed halfedge $h_s$}

        \If{current border ring is not closed}
            \State Find missing edges of current border ring
        \EndIf
        \State Reorder hole vertices
    \EndIf
\EndFor
\end{algorithmic}
\end{algorithm}

As outlined in Algorithm~\ref{alg:alg1}, our approach begins by traversing potential border edges using the half-edge data structure, marking each traversed half-edge. Our focus is on potential border rings initiated from a seed half-edge. Each half-edge, upon traversal, is paired with its preceding half-edge to form a triangle, \(\alpha\). This triangle \(\alpha\) is projected onto a plane defined by the incident faces of the edge's vertices, where it undergoes intersection tests with these faces. 
\\
\begin{wrapfigure}[12]{r}{0.5\columnwidth} 
  \includegraphics[width=0.5\columnwidth]{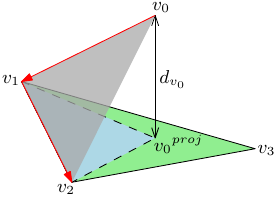} 
  \caption{An example of the intersection test.}
  \label{fig:fig7}
\end{wrapfigure}
Figure~\ref{fig:fig7} illustrates this process. The seed half-edge \(v_{1}v_{2}\) and its preceding half-edge \(v_{0}v_{1}\), marked in red, constitute the virtual triangle \(\alpha\). Meanwhile, triangle \(\beta\) (highlighted in green and defined by \(v_{1}v_{2}v_{3}\)) is an incident face of the vertices of half-edge \(v_{1}v_{2}\). The projection of \(\alpha\) onto \(\beta\) results in an intersecting area, shown in blue. The maximum distance \(D\) from \(\alpha\) to \(\beta\) is equal \(d_{v_0}\) in this example, indicating the distance between vertex \(v_{0}\) and its projected counterpart \(v_{0}^{proj}\). The subsequent equation provides a detailed computation:

\begin{equation}\label{equ:1}
	D = \max(d_{v_0}, d_{v_1}, d_{v_2})
\end{equation}
\begin{equation}\label{equ:2}
	A = \frac{A_{w}}{A_{ \beta }} 
\end{equation}

In Equation~\ref{equ:1}, \(d_{v_0}, d_{v_1}, d_{v_2}\) denote the distances from the three vertices of triangle \(\alpha\) to the plane of the incident triangle face \(\beta\). \(D\) is the maximum of these distances. It should be noted that \(d_{v_1}\) and \(d_{v_2}\), as shown in Figure~\ref{fig:fig7}, can be non-zero when they are identified as duplicated vertices. As indicated in Equation~\ref{equ:2}, \(\alpha\) is then projected onto \(\beta\)'s plane, where \(A_w\) signifies the area of intersection between the projected \(\alpha\) and \(\beta\), \(A_{\beta}\) is the area of \(\beta\), and \(A\) indicates the ratio of these areas. An intersection is deemed valid only when \(D < \epsilon_d = 0.1\) and \(A > \epsilon_t = 0.01\). It should be emphasized that the thresholds for distance and area specified in this paper are measured in meters, and their settings are based on the geometric precision of the test data. In cases where intersections are absent, this pair of edges is incorporated into the border ring collection for the candidate holes.

\begin{figure*}[ht!]
\centering
\subfigure[]{\label{fig:fig8a}\includegraphics[width=0.48\textwidth]{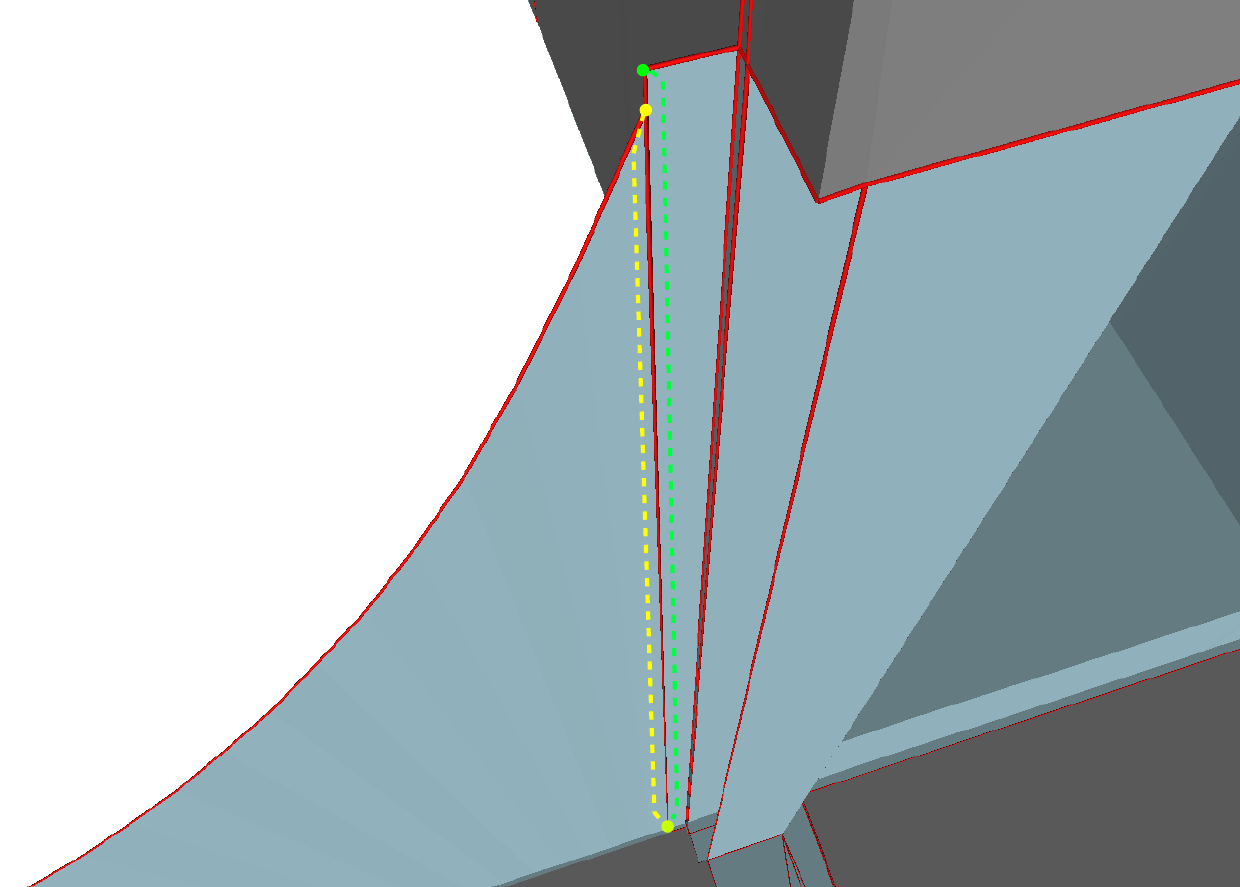}}
\subfigure[]{\label{fig:fig8b}\includegraphics[width=0.48\textwidth]{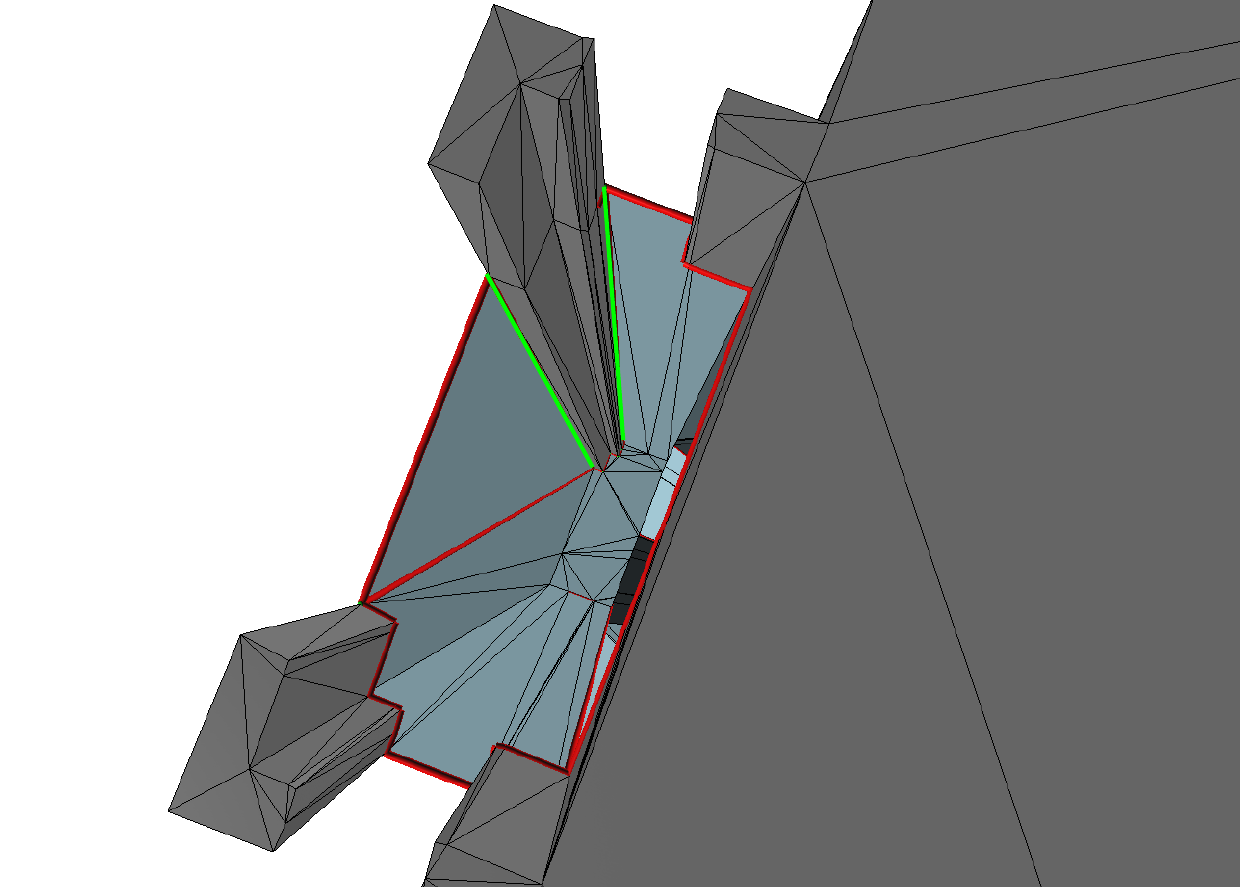}}
\caption{(a) illustrates a hole defined by two overlapping edges that originate from distinct vertices but converge at duplicated vertices, each edge outlined by a yellow dashed curve and a green dashed curve, respectively. (b) illustrates a hole featuring non-manifold edges, highlighted in green. The border of the hole is marked in red, while the mesh's back face is represented in light blue in both images.}
\label{fig:fig8}
\end{figure*}

Next, we conduct closure checks for each border ring, that is, to ascertain whether their start and end points connect. This verification is crucial because meshes plagued with overlapping (see Figure~\ref{fig:fig8a}) or non-manifold edges (see Figure~\ref{fig:fig8b}) can result in border rings that do not close properly. To tackle these issues, we have implemented two strategies. The first addresses overlapping edges: we examine all incident edges of the end vertices of the border ring, including all corresponding duplicated vertices. If an untraversed incident edge is a border edge or overlaps with an edge of the hole, it is added to the border ring collection. The second strategy, concerning non-manifold edges, involves assessing whether all connected non-border ring vertices of the current hole occupy the same plane. If affirmative, the vertex nearest to the border ring's endpoint that meets the intersection test criteria is added to the collection. These strategies are applied in an iterative manner until a fully closed ring is achieved.

Finally, we undertake a reordering of all selected candidate edges. This reordering ensures that adjacent edges are properly linked, creating a continuous and coherent structure. By employing this approach, we ensure that the successfully extracted closed rings of the holes are correctly configured for their integration in the forthcoming stage of remeshing.

\subsection{Remeshing}\label{sec:Footnotes}

The aim of this step is to execute triangle mesh construction, utilizing the closed rings identified in the hole detection process (see Figure~\ref{fig:fig6d}). This is essential for accurately filling in the mesh's missing areas.

We begin by fitting a plane with all the vertices of each identified closed hole ring and project these vertices onto this plane. Utilizing the border edges of the hole as constraints, we then create a 2D constrained Delaunay triangulation from these projected vertices. The subsequent step involves incorporating the triangulated mesh into our input data, aligning it with the corresponding original vertices of the projected points. 

To ensure topological accuracy and avoid issues arising from recovering the 3D structure from the Delaunay triangulation, we conduct degeneracy detection and self-intersection checks for each triangle face to be added. Only faces that meet the specified conditions are ultimately included. Thus, through remeshing, we have obtained a complete building mesh model.

\section{Experiment}\label{sec:experiment}

\subsection{Implementation details}\label{sec:imple}
Our testing environment is powered by a computer configured with an Intel Core i7-7700HQ CPU, operating at a base frequency of 2.8GHz, and bolstered by 32GB of RAM.

We developed our algorithm in C++, leveraging the capabilities of the Easy3D~\cite{nan2021easy3d} library. This library facilitated our handling of triangular mesh read/write operations, manipulation of mesh data structures, and enabled efficient topological detection and validation. For the implementation of constrained Delaunay triangulation and edge-stitching processes, we employed the robust CGAL 5.6 library~\cite{fabri2000design}.

In our comparative experimental assessments, we selected six characteristic models, each featuring complex holes, from The Hague's maunally created CityGML LoD2 building models~\cite{denhaag3d}. These models underwent a conversion process from CityGML (polygonal surface meshes) to OBJ format (triangular meshes) using FME software, specifically for conducting our advanced hole-filling operations.

\subsection{Comparisons}\label{sec:comp}

\begin{figure*}[ht!] %
\begin{adjustwidth}{-3cm}{-3cm}
	\centering
	\begin{tabular}{cccccc}	
		\includegraphics[width=0.16\textwidth]{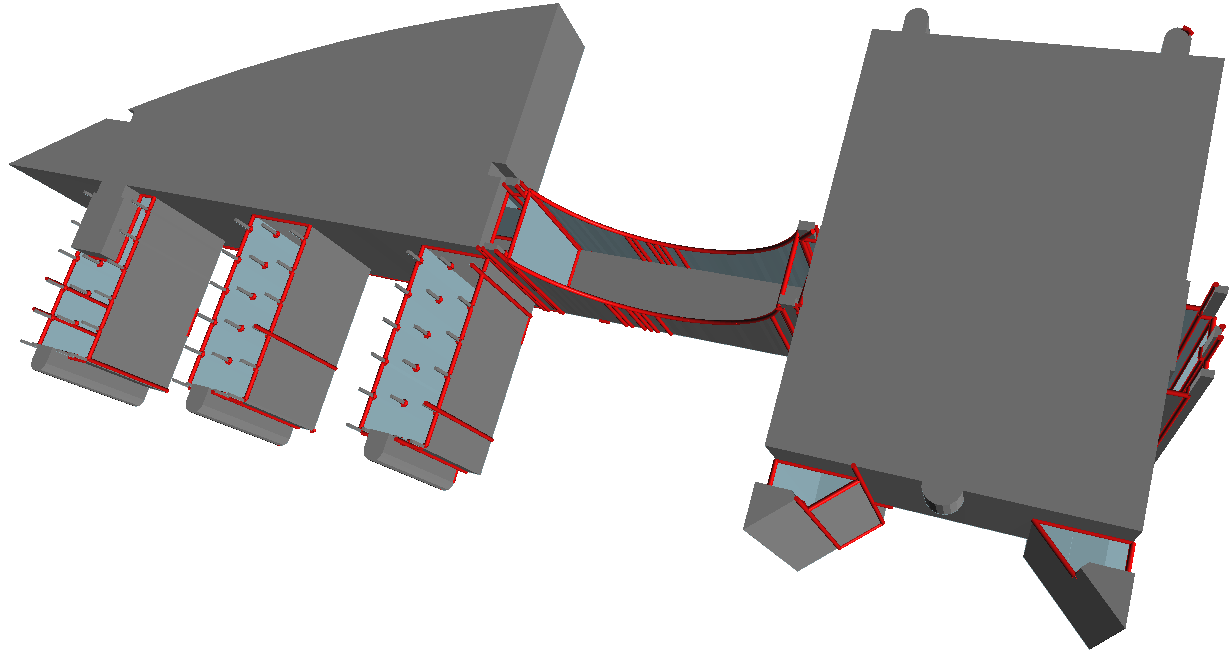}&
		\includegraphics[width=0.16\textwidth]{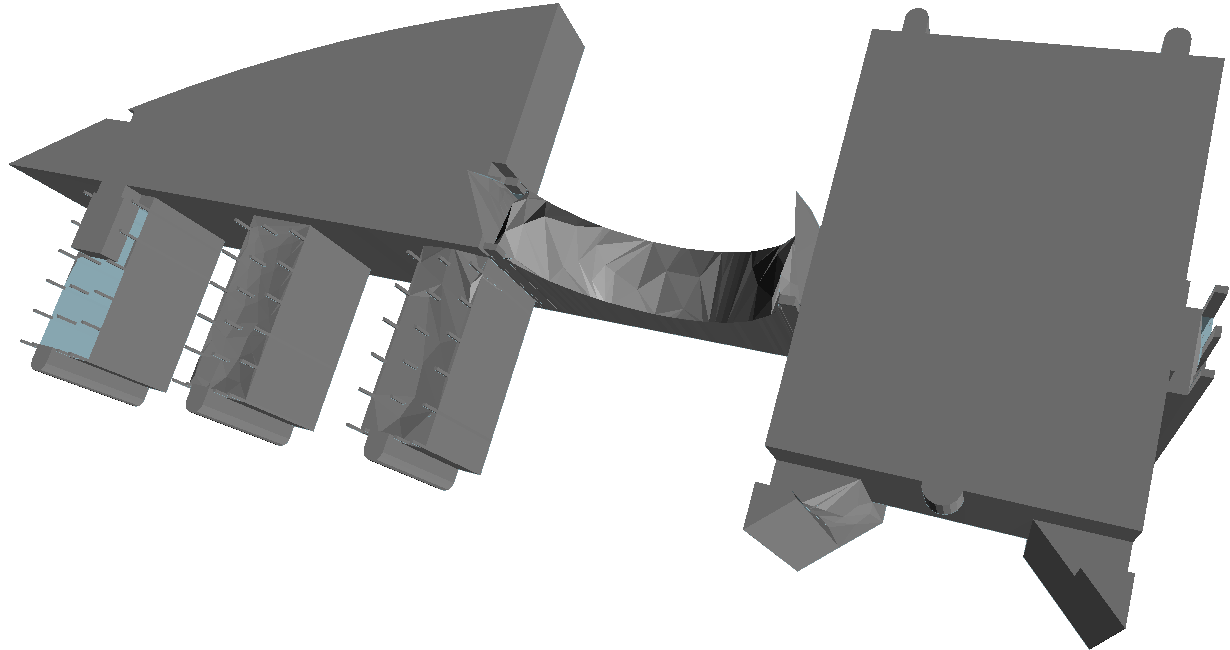}&
		\includegraphics[width=0.16\textwidth]{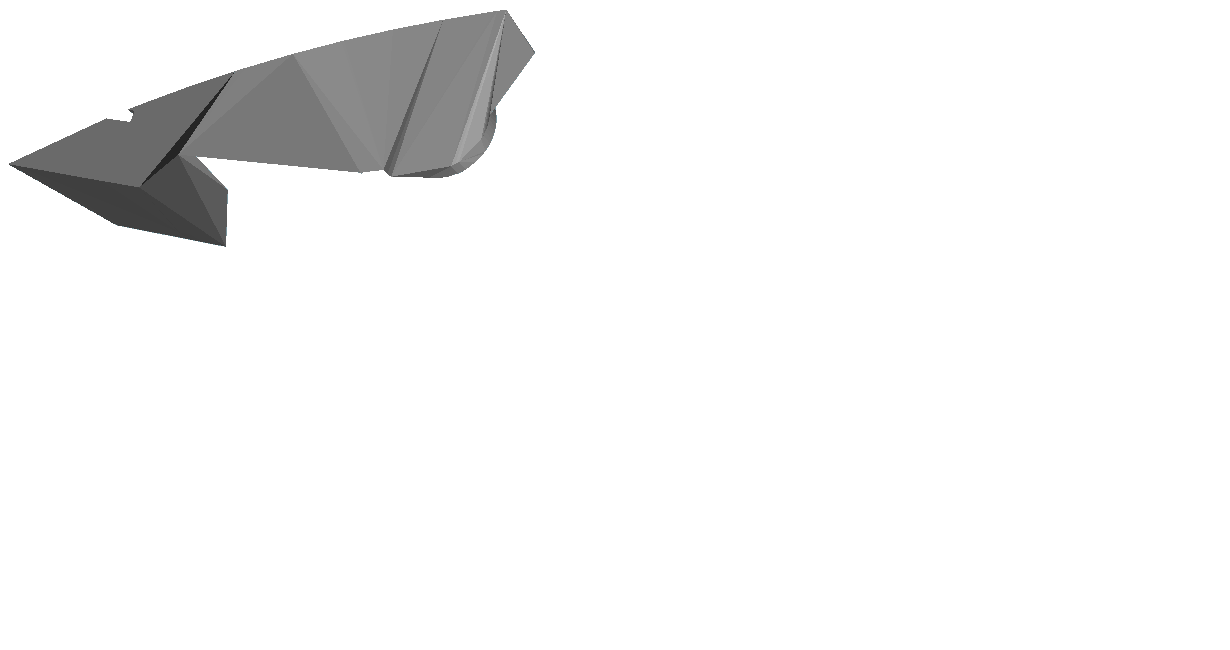}&
		\includegraphics[width=0.16\textwidth]{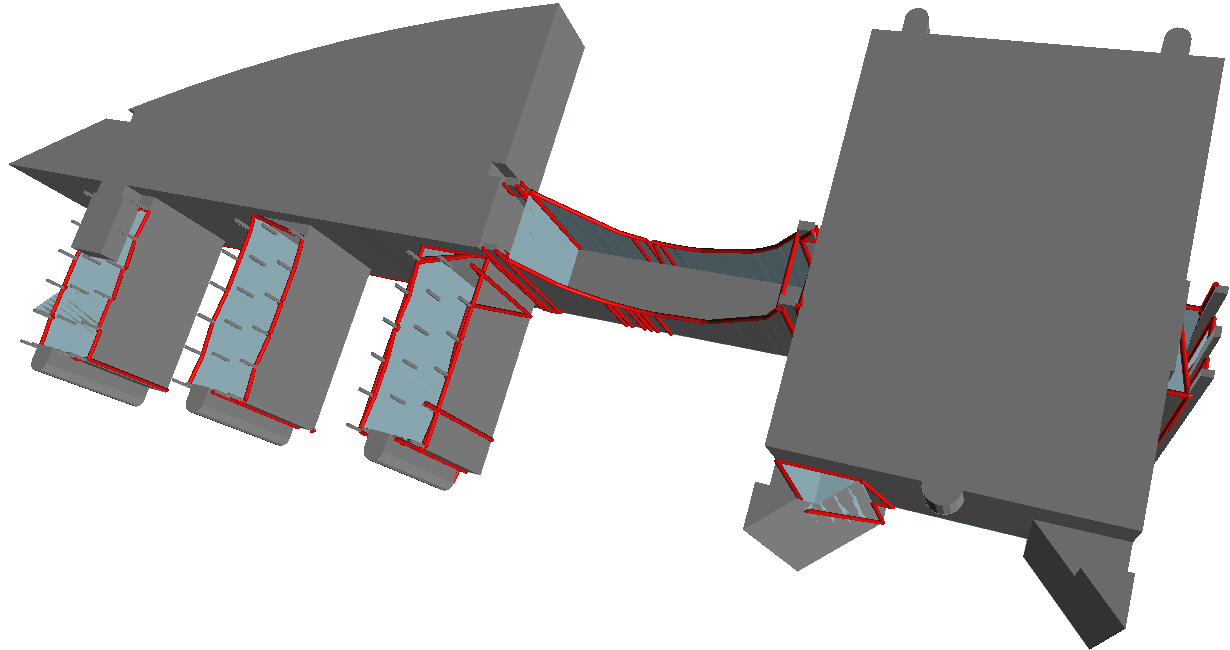}&
  		\includegraphics[width=0.16\textwidth]{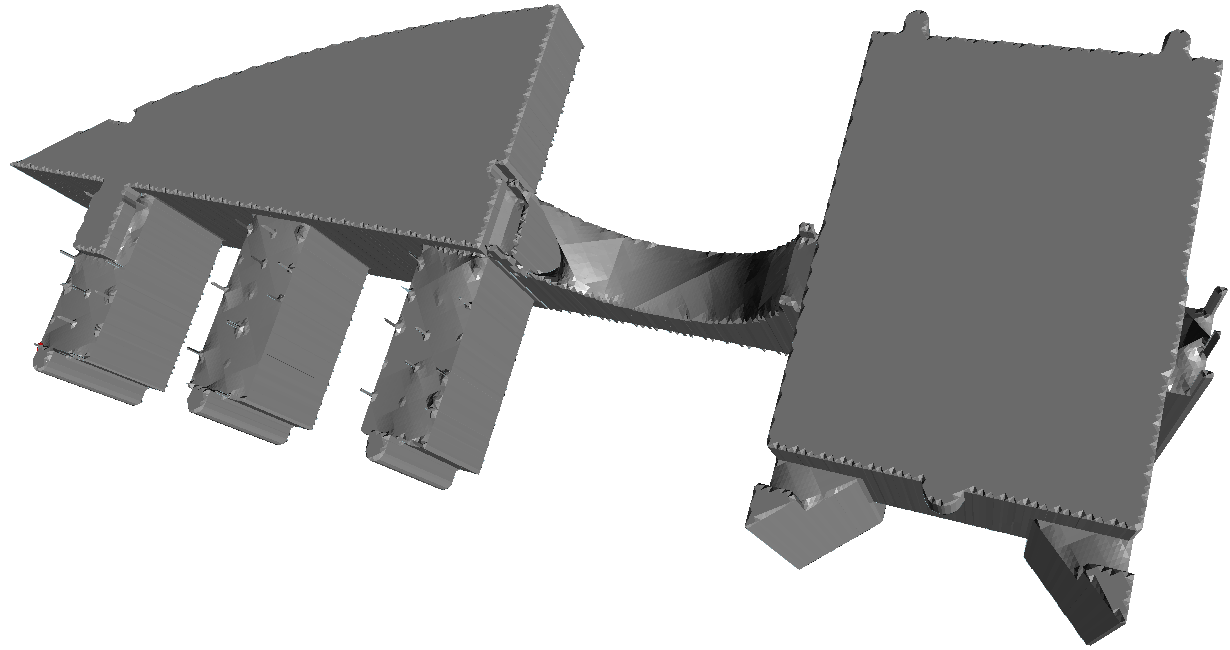}&
		\includegraphics[width=0.16\textwidth]{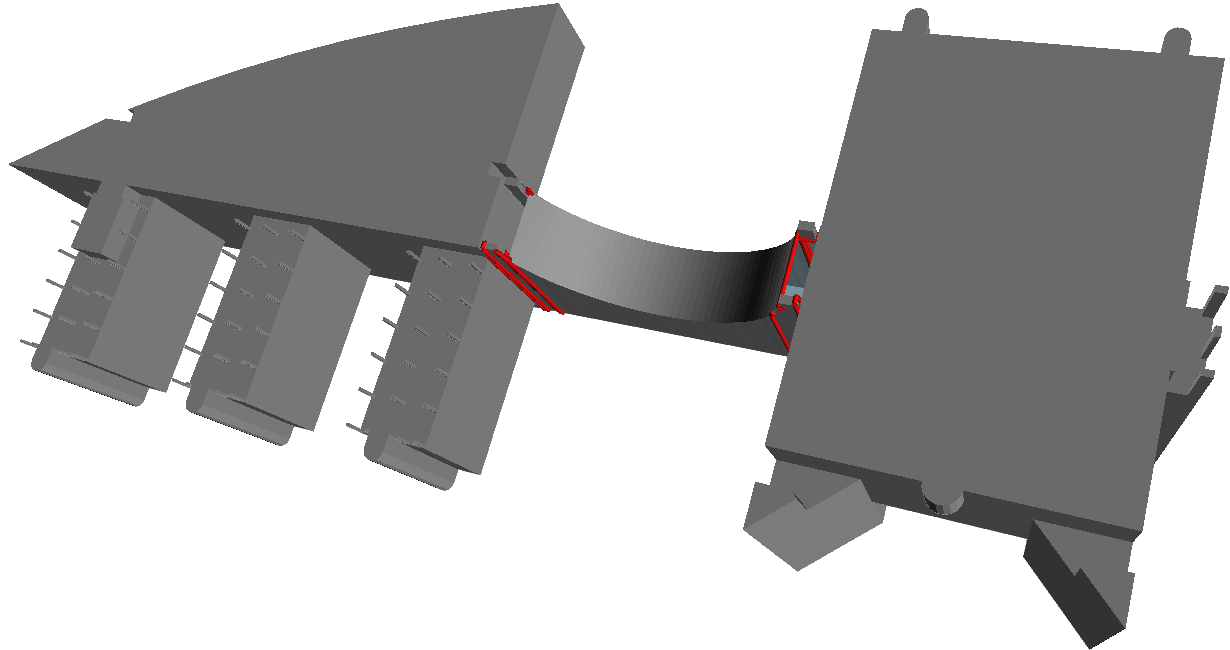}\\
		\midrule	
  		\includegraphics[width=0.16\textwidth]{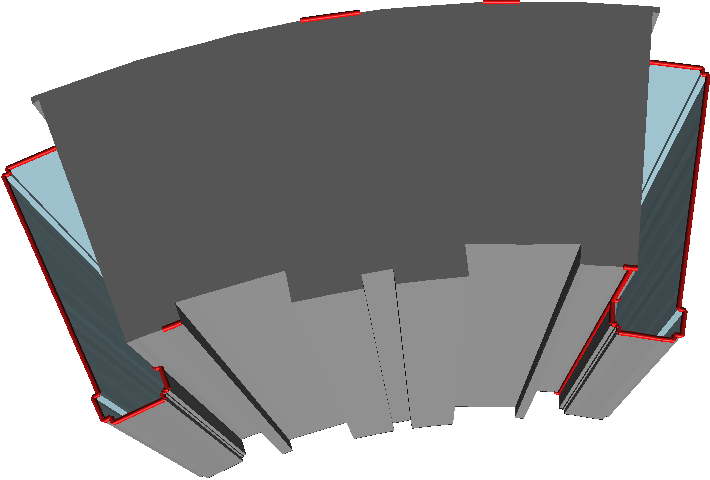}&
		\includegraphics[width=0.16\textwidth]{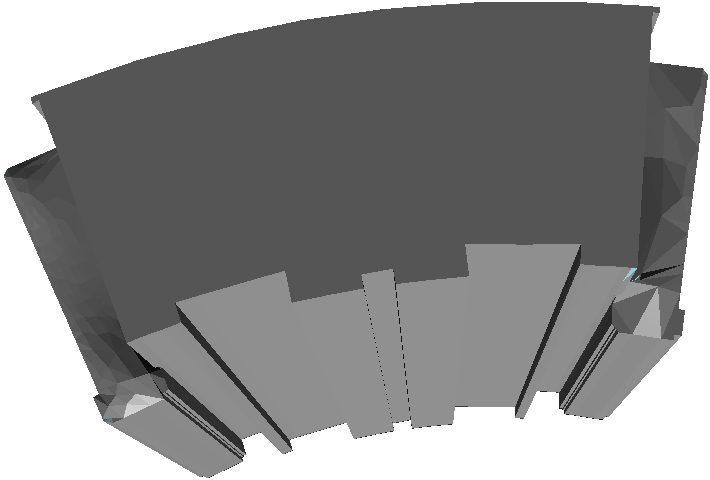}&
		\includegraphics[width=0.16\textwidth]{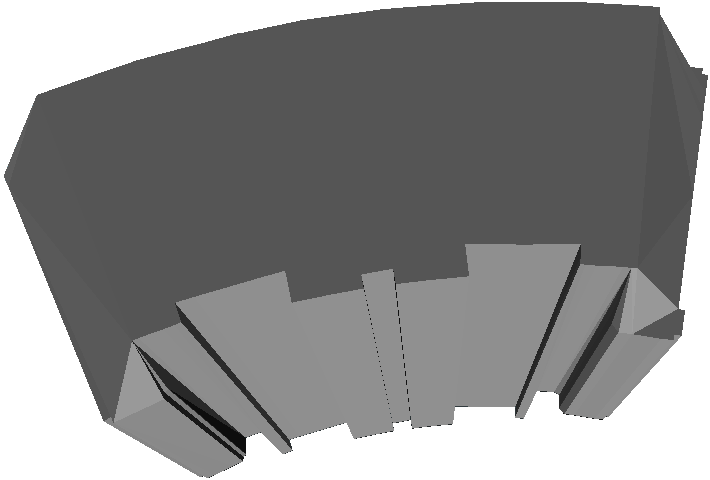}&
		\includegraphics[width=0.16\textwidth]{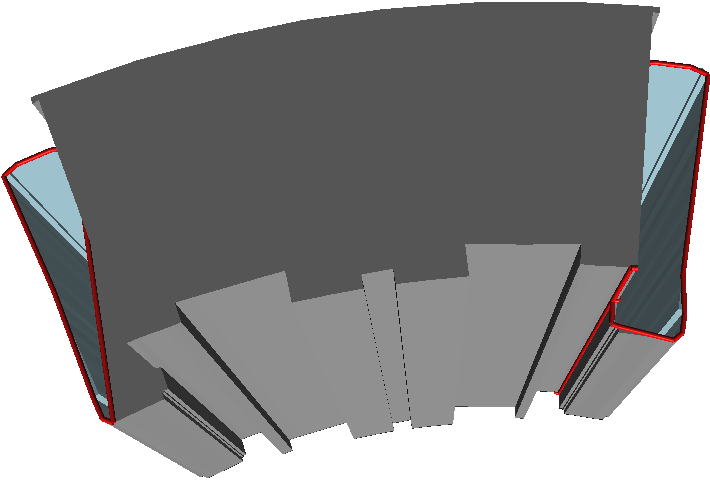}&
  		\includegraphics[width=0.16\textwidth]{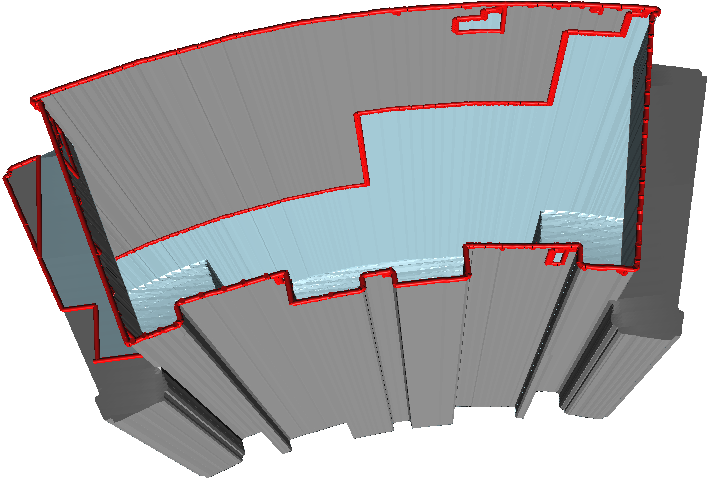}&
		\includegraphics[width=0.16\textwidth]{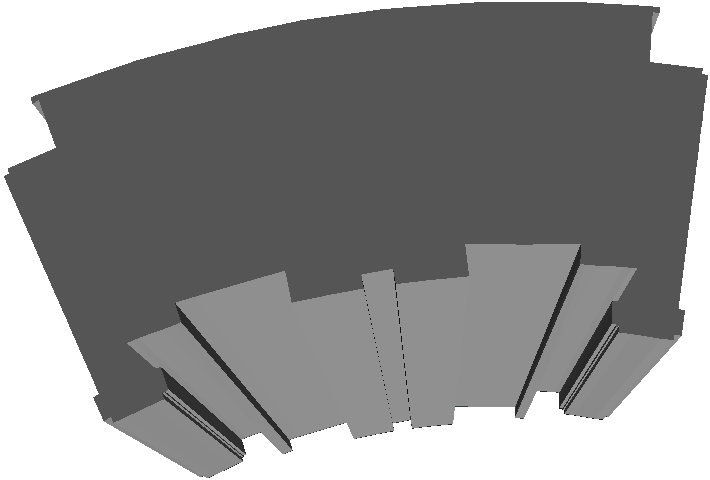}\\
  		\midrule	
  		\includegraphics[width=0.16\textwidth]{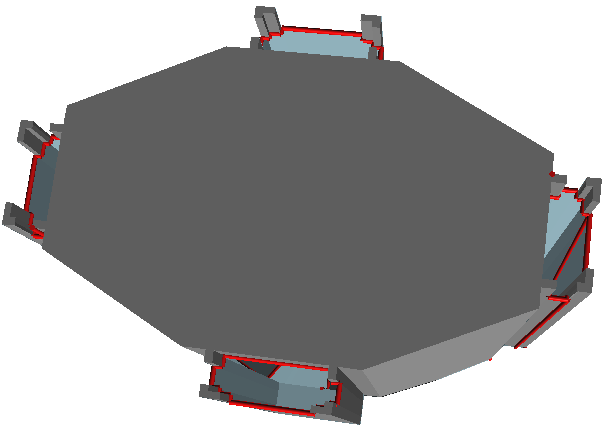}&
		\includegraphics[width=0.16\textwidth]{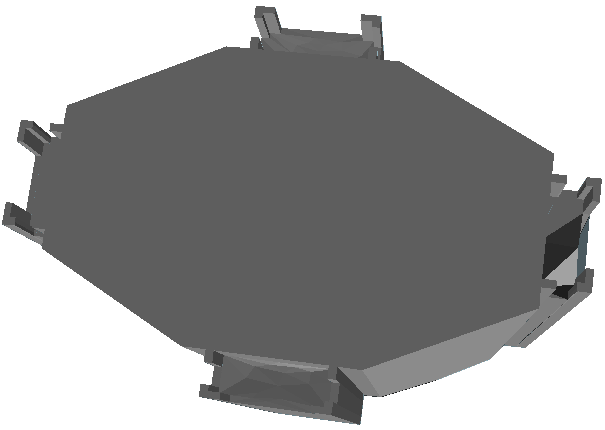}&
		\includegraphics[width=0.16\textwidth]{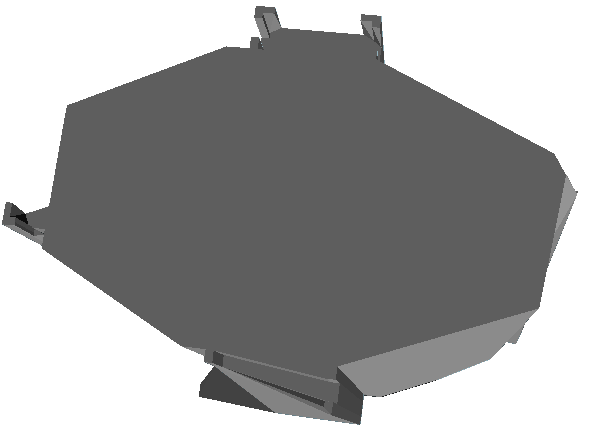}&
		\includegraphics[width=0.16\textwidth]{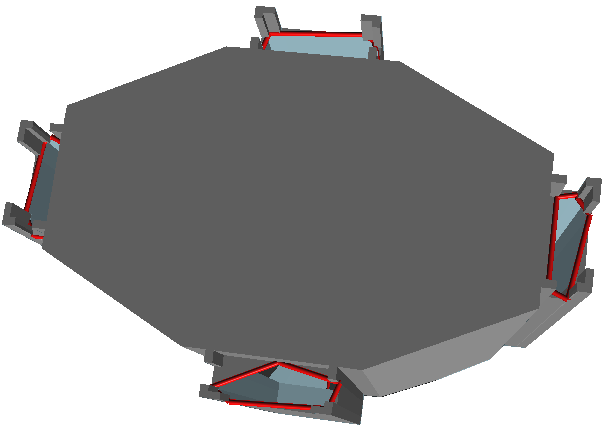}&
  		\includegraphics[width=0.16\textwidth]{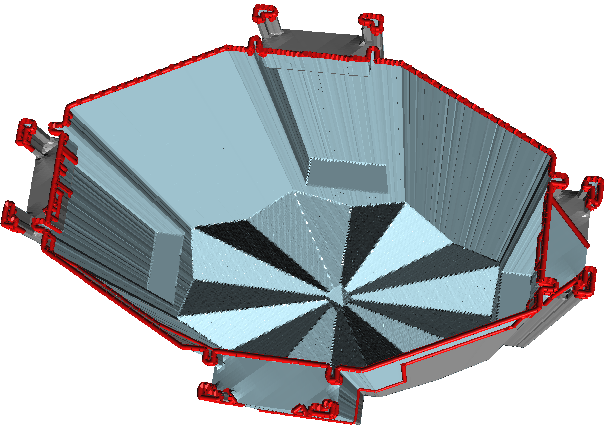}&
		\includegraphics[width=0.16\textwidth]{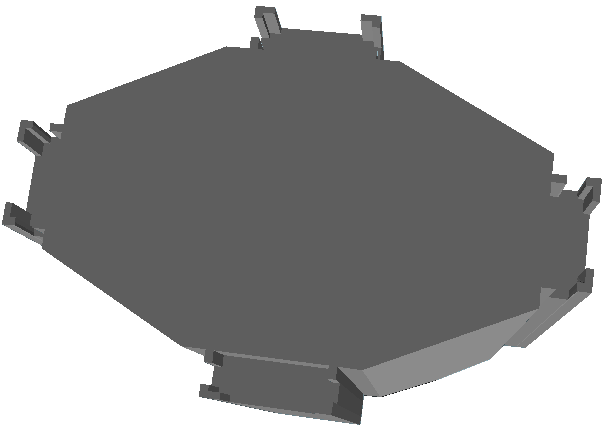}\\
  		\midrule	
  		\includegraphics[width=0.16\textwidth]{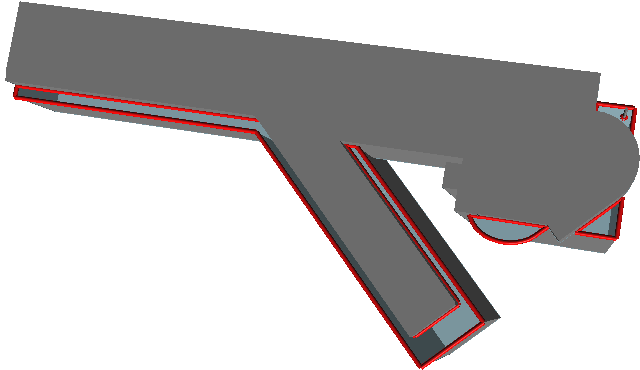}&
		\includegraphics[width=0.16\textwidth]{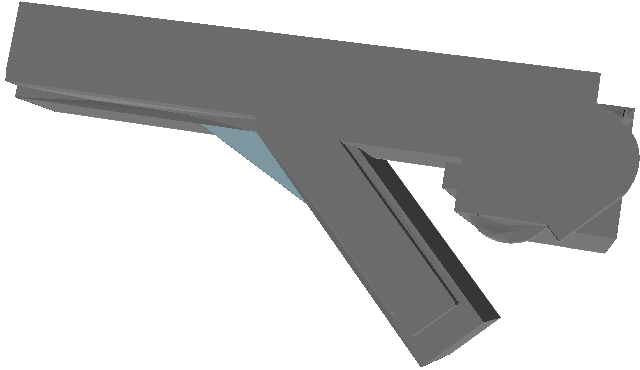}&
		\includegraphics[width=0.16\textwidth]{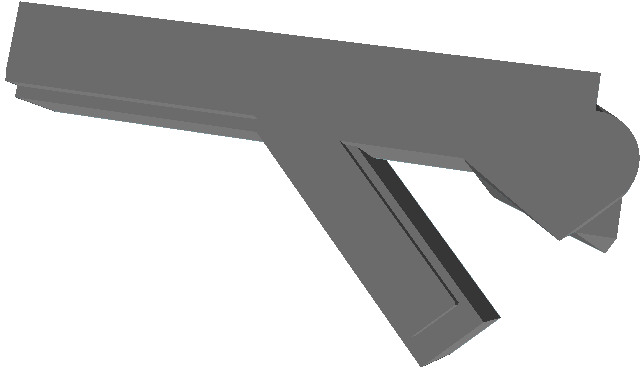}&
		\includegraphics[width=0.16\textwidth]{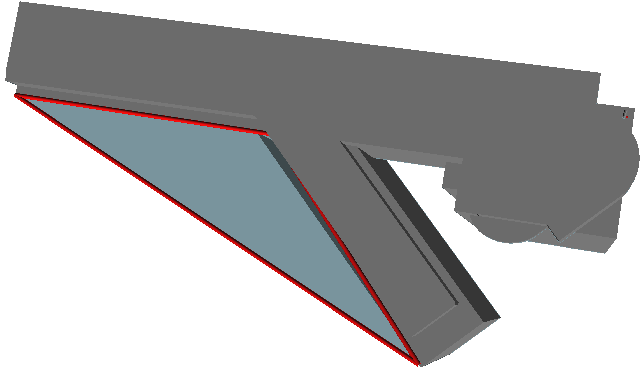}&
  		\includegraphics[width=0.16\textwidth]{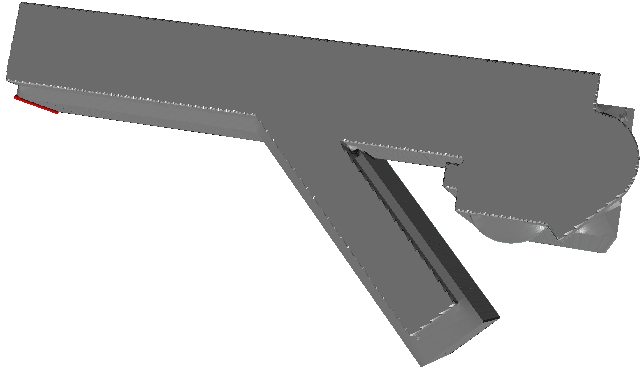}&
		\includegraphics[width=0.16\textwidth]{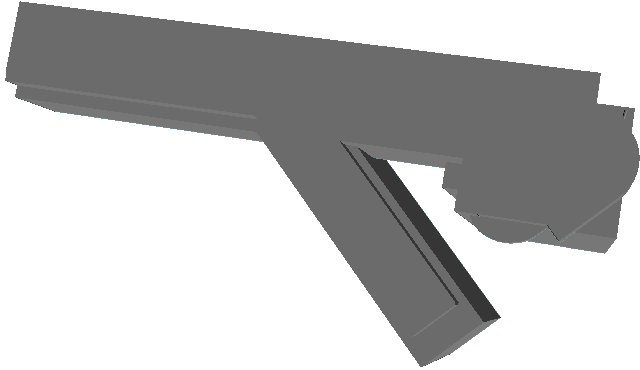}\\
  		\midrule	
  		\includegraphics[width=0.16\textwidth]{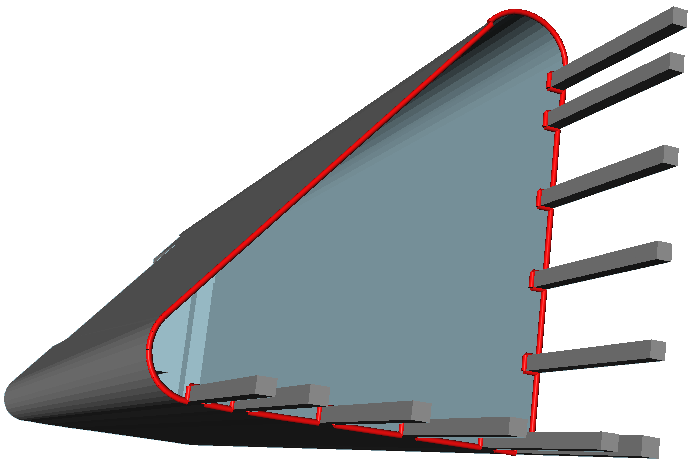}&
		\includegraphics[width=0.16\textwidth]{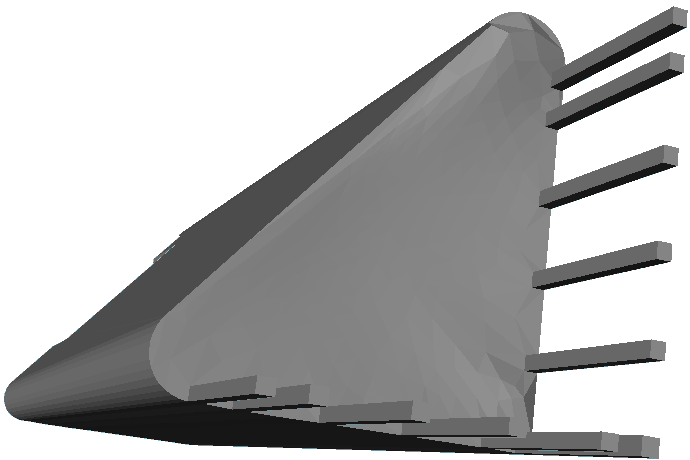}&
		\includegraphics[width=0.16\textwidth]{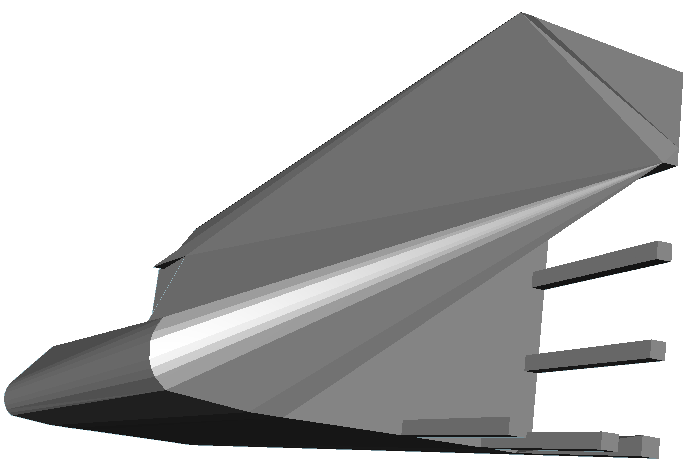}&
		\includegraphics[width=0.16\textwidth]{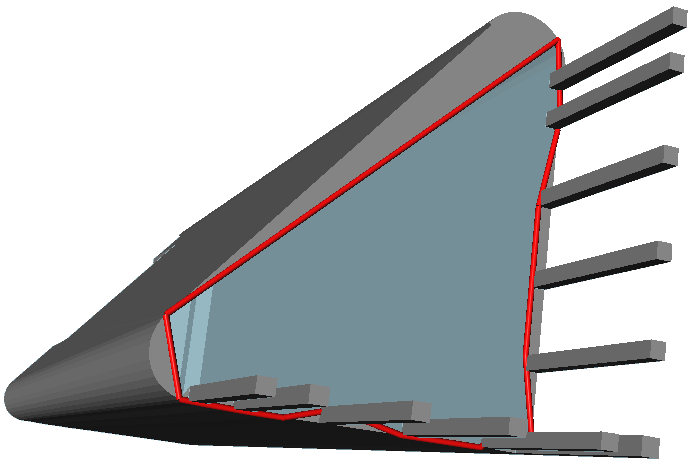}&
  		\includegraphics[width=0.16\textwidth]{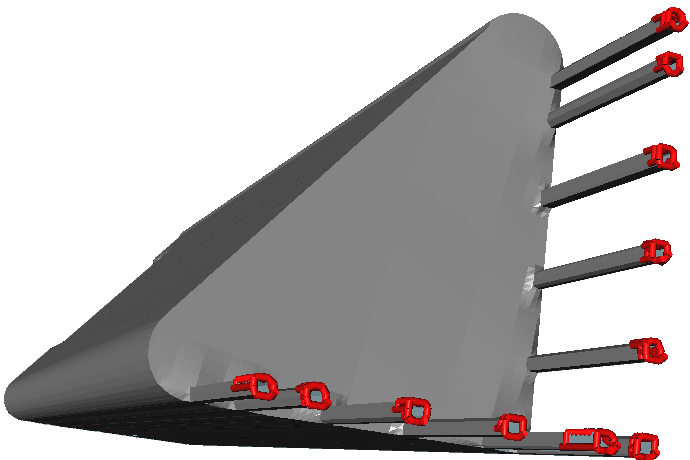}&
		\includegraphics[width=0.16\textwidth]{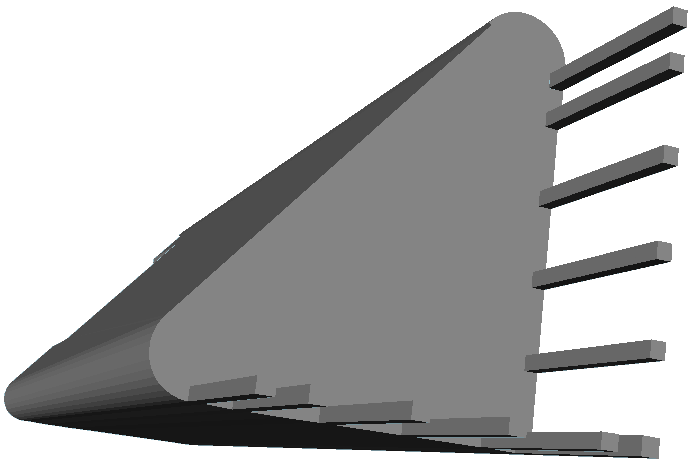}\\
  		\midrule	
  		\includegraphics[width=0.16\textwidth]{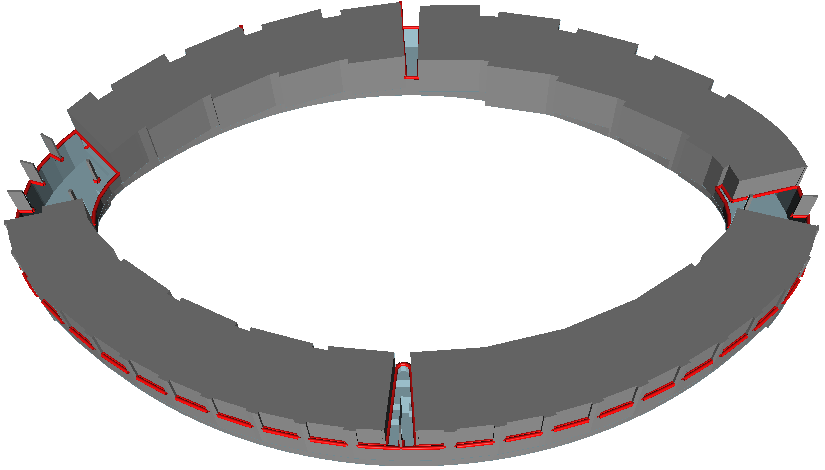}&
		\includegraphics[width=0.16\textwidth]{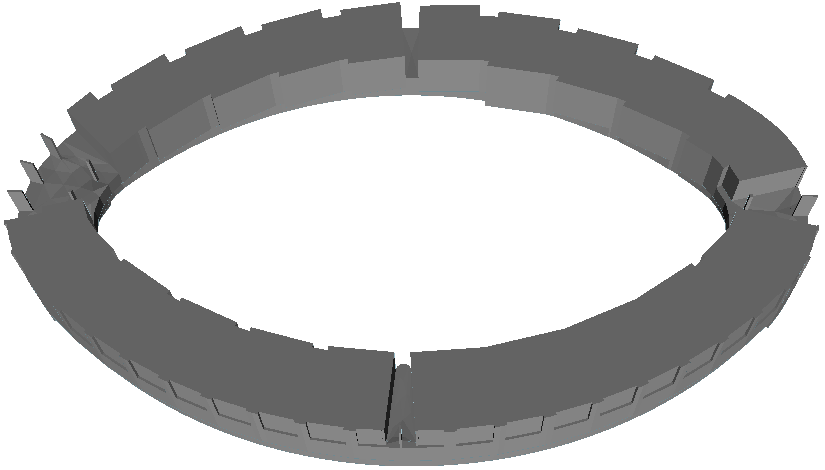}&
		\includegraphics[width=0.16\textwidth]{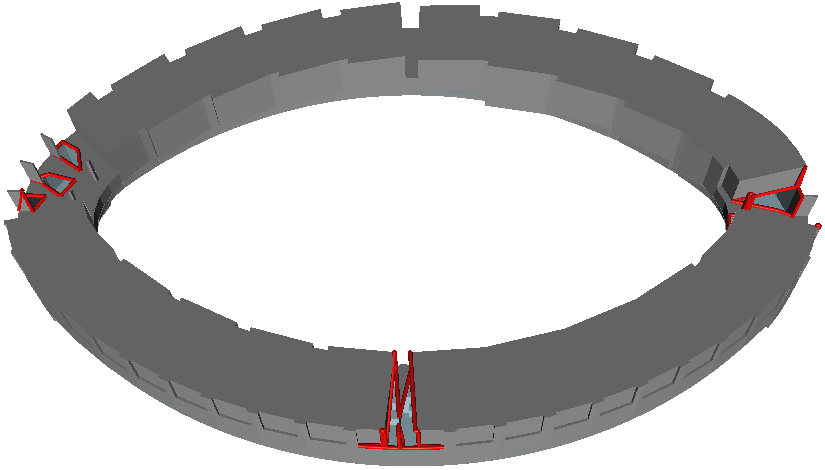}&
		\includegraphics[width=0.16\textwidth]{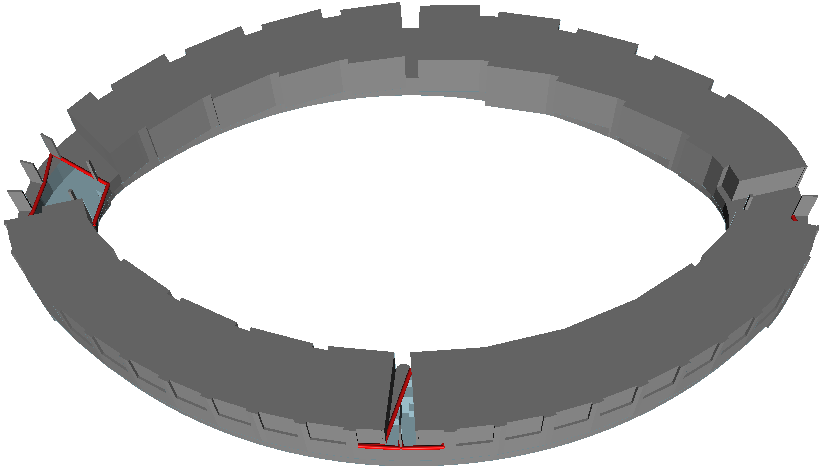}&
  		\includegraphics[width=0.16\textwidth]{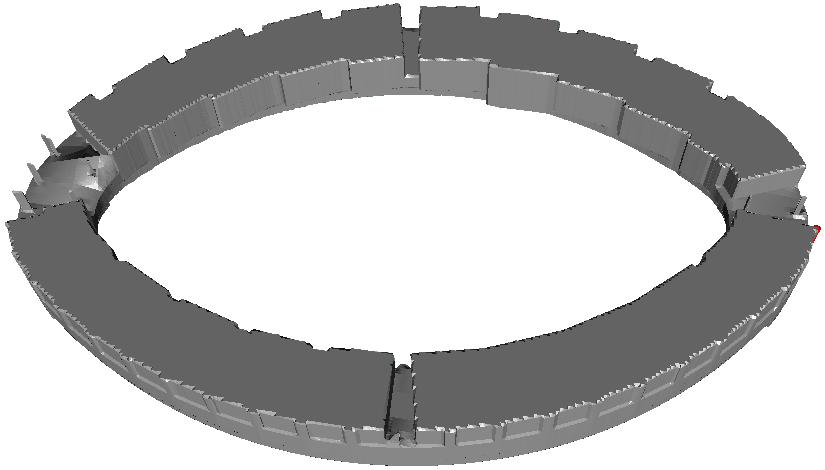}&
		\includegraphics[width=0.16\textwidth]{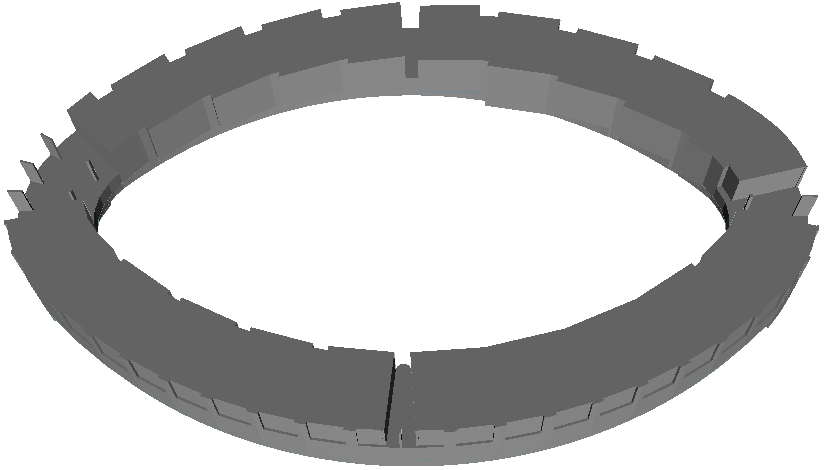}\\
		(a) Input&
		(b) PMP&
		(c) MeshFix&
		(d) MeshLab&
            (e) Polymender&
		(f) Ours\\	
	\end{tabular}
\end{adjustwidth}
\caption{Qualitative comparisons of hole-filling methods in building mesh models. The boundaries of the holes are highlighted in red, while the back face of the mesh is depicted in light blue. }
\label{fig:fig9}
\end{figure*}

\begin{table}[h]
\centering
\begin{tabular}{@{}lccccc@{}}
\toprule
            & MeshL. & PMP & MeshF. & Polym. & Ours     \\ \midrule
Time (s)    & 0.76   & 2.12 & 1.24   & 1.13     & 16.19 \\
Face num. & 12.3k   & 16.2k  & 7k    & 158k    & 15.5k   \\ \bottomrule
\end{tabular}
\caption{Comparative analysis of hole filling methods: Evaluating time efficiency and total output face count (Face Num.) across six chosen building meshes (incorporating 11.3k input faces) utilizing MeshLab (MeshL.), PMP, MeshFix (MeshF.), PolyMender (PolyM.), and our method. }
\label{tab:tab1}
\end{table}

We carried out a comprehensive qualitative analysis and a performance comparison between our mesh hole-filling method and other established methods, such as PMP~\cite{pmp-library}, MeshFix~\cite{attene2010lightweight}, MeshLab~\cite{cignoni2008meshlab}, and Polymender~\cite{ju2004robust}. The experimental results demonstrate that our method excels in completing LoD2 building models, surpassing all traditional methods on the quality of the output mesh, at the cost of an admittedly longer running time.

As illustrated in Figure~\ref{fig:fig9}, the PMP method, while successfully filling all holes, but generates erroneous surfaces due to the presence of pseudo-holes (as highlighted in rows 1 and 4 of the second column in Figure~\ref{fig:fig9}). Additionally, PMP's surface fairing post-processing step fails to ensure completely flat surfaces. 
Conversely, MeshFix alters the original geometric structure of the input data because of non-manifolds, pseudo-holes, and disconnected components, resulting in the loss of specific geometric structures in its results. In contrast, MeshLab's hole-filling approach maintains the geometric structure of the initial data but falls short in terms of completeness, leaving some holes unfilled. Polymender, employing a reconstruction-based algorithm, over-smooths sharp geometric features of the original input data and in some cases introduces new holes (as depicted in rows 2 and 3 of the fifth column in Figure~\ref{fig:fig9}). Our method distinguishes itself by preserving the uniformity of local geometry at the holes and by upholding structural completeness. Moreover, our method showcases remarkable adaptability in processing input data with topological inaccuracies, while also providing improved stability in its operation.

Table~\ref{tab:tab1} reveals that our algorithm's running time is longer than other methods, largely due to the additional time required for addressing topological errors such as non-manifolds, duplicate vertices, and overlapping edges. Notably, given a total number of 11.3k triangle faces in the input data, the increased number of triangles with our method is less than PMP. This benefit primarily stems from our method's ability to discern between true and pseudo-holes. Significantly, Polymender, a method based on reconstruction, produces an excessively large number of triangles to minimize geometric differences with the original data, which runs counter to the objective of preserving the LoD2 building model's simplicity. MeshFix, on the other hand, results in a loss of too many triangles, yielding a more incomplete final result.

To conclude, despite a longer running time, our method excels over others in terms of delivering complete and geometrically uniform hole-filling results.

\subsection{Limitations and future prospects}\label{sec:imple}

\begin{figure}[ht!]
\begin{center}
        \includegraphics[width=0.48\columnwidth]{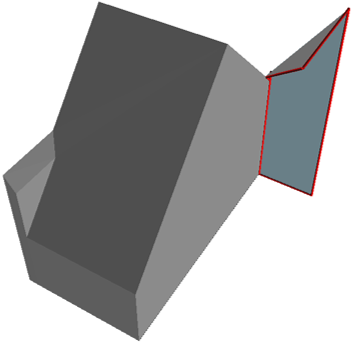}
        \includegraphics[width=0.48\columnwidth]{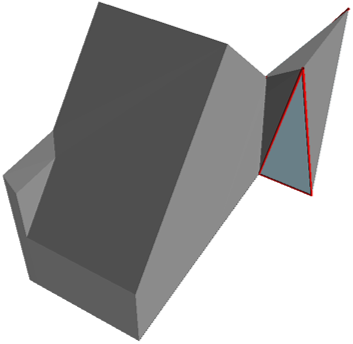}
        \includegraphics[width=0.48\columnwidth]{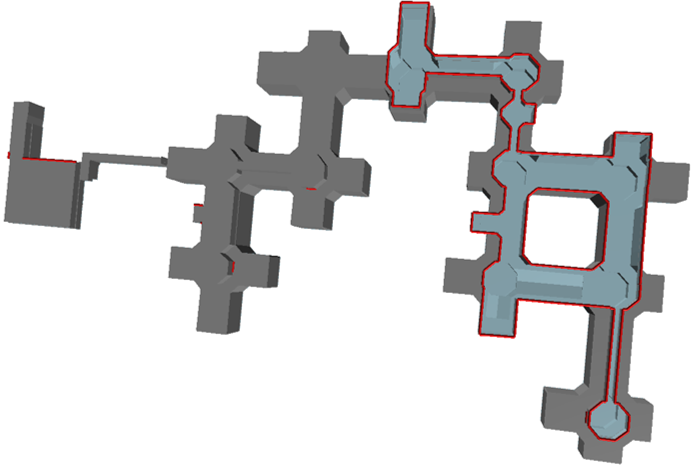}
        \includegraphics[width=0.48\columnwidth]{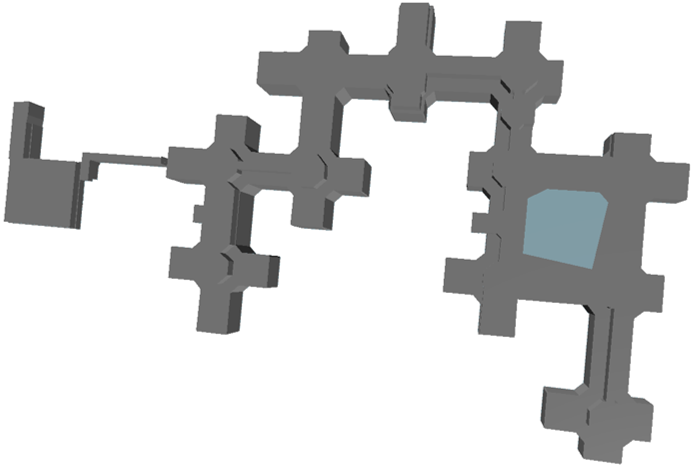}
	\caption{Failure cases of our algorithm. The boundaries of the holes are highlighted in red, while the back face of the mesh is depicted in light blue.}
\label{fig:fig10}
\end{center}
\end{figure}

While our algorithm effectively addresses most topological errors in the input data and reconstructs the complete geometric surface of building mesh holes, it encounters limitations in three specific scenarios:
Firstly, in cases where the hole of the input model has significantly lost geometric structure — a situation exemplified in the upper part of Figure~\ref{fig:fig10} — where only a few geometric planar structures are near the hole, our algorithm is unable to fully reconstruct the hole's geometric surface. Secondly, if the genus of the closed original input mesh model after completion exceeds zero, and the location of a detected geometric hole to be filled coincides with the genus hole, our algorithm might inadvertently produce faces to close the genus hole, as illustrated in the bottom part of Figure~\ref{fig:fig10}. Thirdly, when the input model has disconnected components and a hole boundary is split between the two components, our algorithm struggles to form a complete and closed ring around the hole (see Figure~\ref{fig:fig9} (f) top). 
Notice that for the above three situations, the other comparative algorithms mentioned in Section~\ref{sec:comp} also fail. Hence, solving these problems would be an interesting future research direction.

Moreover, for flawed LoD2 building models, resolving only the hole issue may not be adequate for downstream application requirements. This is because, apart from holes, the presence of issues like misoriented normals and inner faces can also greatly impact the overall quality of the final model. Consequently, the comprehensive goal in repairing LoD2 building models is to achieve a watertight, 2-manifold model with correctly oriented normals. We will take this goal as our future research direction.

\section{Conclusions}\label{sec:conclusion}
In conclusion, our research contributes a novel algorithm to the field of urban model repair, particularly for filling holes in LoD2 building models. Our method surpassing traditional techniques in terms of preserving the original geometric structure and handling various topological errors. The algorithm's effectiveness is underscored by its ability to maintain geometric structure uniformity and integrity, proving advantageous over existing methods in qualitative and performance comparisons. While our approach represents a significant advancement, it also encounters certain limitations, particularly in addressing severely missing geometric structures, models with overlapping holes and non-zero genus, and entirely disconnected components. These challenges, alongside the broader goal of achieving comprehensive repair for LoD2 building models, set the direction for future research. The ultimate aim is to develop techniques that not only address holes but also rectify other critical aspects such as misoriented normals and inner faces, moving towards creating watertight, 2-manifold models with correctly oriented normals, thereby enhancing the utility and applicability of 3D urban building models.

{
	\begin{spacing}{1.17}
		\normalsize
		\bibliography{Filling_holes_in_LoD2_building_models} 
	\end{spacing}
}

\end{document}